\documentclass[aps,prl,twocolumn,superscriptaddress]{revtex4-2}

\usepackage{graphicx}
\usepackage{dcolumn}
\usepackage{bm}

\begin{document}

\preprint{APS/123-QED}

\title{The Counterintuitive ‘One-Enhanced, One-Suppressed’ Bifurcation Phenomenon of Spin Mobility under Weak Spin-Orbit Coupling in Helices}%

\author{Hengrui Yang}
  \affiliation{%
 MOE Key Laboratory of Organic Optoelectronics and Molecular Engineering, Department of Chemistry, Tsinghua University, Beijing 100084, China%
}
\author{Yu Xiong}%
  \affiliation{%
 MOE Key Laboratory of Organic Optoelectronics and Molecular Engineering, Department of Chemistry, Tsinghua University, Beijing 100084, China%
}
\author{Weitang Li}%
   \affiliation{%
 Guangdong Basic Research Center of Excellence for Aggregate Science, School of Science and Engineering, The Chinese University of Hong Kong, Shenzhen, Guangdong 518172, China
}
\author{Zhigang Shuai}%
  \email{shuaizhigang@cuhk.edu.cn}
  \affiliation{%
 MOE Key Laboratory of Organic Optoelectronics and Molecular Engineering, Department of Chemistry, Tsinghua University, Beijing 100084, China%
}%
  \affiliation{%
 Guangdong Basic Research Center of Excellence for Aggregate Science, School of Science and Engineering, The Chinese University of Hong Kong, Shenzhen, Guangdong 518172, China
}


\begin{abstract}
The chiral-induced spin selectivity (CISS) effect poses a longstanding puzzle: how weak spin-orbit coupling (SOC) within a helical molecule produces strong spin polarization. We uncover a counterintuitive phenomenon, a ”one-enhanced, one-suppressed” bifurcation of spin-resolved mobility relative to the SOC-free case. We trace this phenomenon to an analytical sign reversal of spin-dependent transmission and identify it as the microscopic origin of CISS. The phenomenon is found absent in achiral systems and single-stranded DNA, validating our theoretical picture. Our findings resolve the core puzzle and establish the origin of CISS.
\end{abstract}

\maketitle

\emph{Introduction} --- The chiral-induced spin selectivity (CISS) effect, which enables chiral molecules to filter electron spins without magnetic components, has been well documented across a wide range of experiments\cite{ray_asymmetric_1999,gohler_spin_2011,mishra_spin-dependent_2013,kim_chiral-induced_2021}. However, its origin remains controversial\cite{naaman_chiral_2019,guo_spin-selective_2012,alwan_spinterface_2021}.  A central question in CISS concerns how the weak spin-orbit coupling (SOC) present in light atoms\cite{kuemmeth_coupling_2008} can still produce substantial spin polarization\cite{gohler_spin_2011}. 

A variety of theoretical mechanisms have been proposed, including scattering in helical potentials\cite{yeganeh_chiral_2009,medina_chiral_2012,varela_inelastic_2013,eremko_spin_2013,gersten_induced_2013}, quantum transport\cite{guo_spin-selective_2012,gutierrez_spin-selective_2012,guo_spin-dependent_2014,guo_spin_2016,dalum_theory_2019,sarkar_spin-selective_2019}, many-body effects\cite{fransson_vibrational_2020,fransson_chirality-induced_2019,fransson_charge_2021,zhang_highly_2023,das_temperature-dependent_2022,chiesa_many-body_2024}, polaron formation\cite{zhang_chiral-induced_2020}, Berry phase\cite{teh_spin_2022,liu_enhancement_2024}, and spinterface effects\cite{alwan_spinterface_2021,alwan_role_2024}. Not all of these, however, directly address the central question of what underlies CISS. Some circumvent it by simply not discussing the role of SOC in generating polarization. Others introduce unrealistically large SOC strength\cite{guo_spin-selective_2012} or Zeeman-like fields\cite{zhang_chiral-induced_2020,kato_electronic_2023,baranov_tunneling_2026,sun_quantum_2005} in their models. Many incorporate electrodes, especially ferromagnetic electrodes\cite{fransson_vibrational_2020,fransson_charge_2021,fransson_chirality-induced_2019,das_temperature-dependent_2022,zhang_highly_2023} in their calculations, motivated by experimental setups\cite{gohler_spin_2011,mishra_spin-dependent_2013,naaman_chiral_2020,carmeli_spin_2014,kumar_device_2013,zhang_highly_2023,das_spin-induced_2023}. And some, reasoning that molecular SOC alone is too weak, directly attribute the polarization to the metal electrodes themselves\cite{alwan_spinterface_2021,alwan_role_2024}. Yet the core question, whether, how and why such weak SOC in the molecule can generate substantial polarization, remains unresolved. Conventional intuition would suggest that it should produce at most a tiny correction to both spin channels.

\begin{figure}[b]
\includegraphics[width = 1.0\columnwidth]{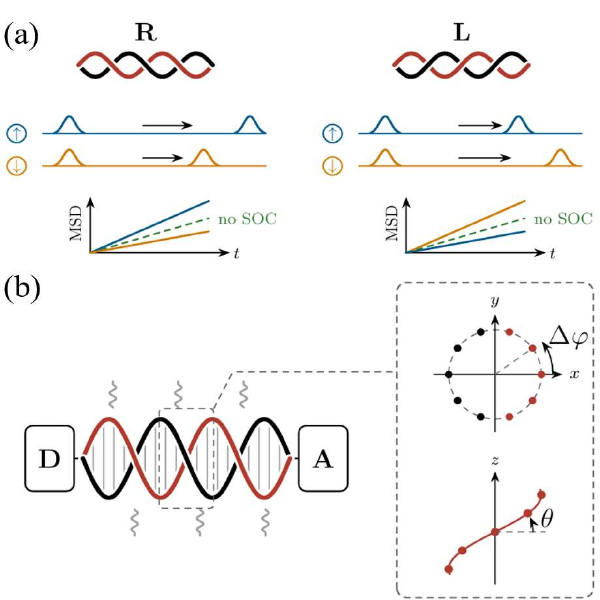}
\caption{\label{fig:sketch} (a) Illustration of the "one-enhanced, one-suppressed" bifurcation. The spin-resolved mean-square displacement (MSD) curves for the right-handed helix, with and without SOC, are shown for comparison. (b) Schematic representation of the donor–dsDNA–acceptor system, with the helix characterized by $\theta$ and $\Delta\varphi$. }
\end{figure}

In this Letter, we uncover a counterintuitive bifurcation, a new physical phenomenon that constitutes the most fundamental question of CISS. Relative to the SOC-free baseline, weak SOC does not merely produce a tiny correction. Instead, one spin mobility is enhanced while the other is suppressed in absolute terms, yielding a substantial spin polarization [Fig.~\ref{fig:sketch}(a)]. Flipping the helix handedness reverses the enhanced and suppressed spin channels. We prove analytically that the bifurcation originates from opposite signs of the linear-order SOC corrections to the two spin transmission probabilities. We demonstrate this phenomenon in a minimal donor–helix molecule–acceptor transport model (exemplified by double-stranded DNA, dsDNA), as illustrated in Fig.~\ref{fig:sketch}(b). In this model, spin-independent dissipation locks the SOC-induced asymmetry into a steady-state bifurcation. Control results on non-helical duplexes or single-stranded DNA (ssDNA) confirm that SOC must be artificially increased by orders of magnitude to produce any detectable but symmetric mobility change, validating our theoretical picture.


Our findings thus establish the “one-enhanced, one-suppressed” bifurcation not only as a new physical phenomenon in its own right, but also as the microscopic origin that resolves the core puzzle of how why weak SOC within helical molecule can generate strong spin polarization.

\emph{General Setup} --- We consider a donor–helix molecule–acceptor (D-M-A, where D, M and A denote donor, molecular bridge and acceptor, respectively) transport model in which a helical molecular bridge (exemplified by double-stranded DNA, dsDNA) couples donor and acceptor at its two ends\cite{chiesa_many-body_2024,zhang_dynamical_2025}. The helix is modeled as a tight-binding chain of N sites with two strands. The system Hamiltonian is $H_S=H_M+H_{DM}+H_{MA}$. The molecular Hamiltonian $H_{M}$ describes the helical bridge and includes intra‑chain hopping $t_j$, onsite energy $\varepsilon_j$, inter‑chain coupling $\lambda$, and a helical SOC term of strength $t_{\mathrm{SO}}$:
\begin{eqnarray}
H_M&=&\sum_{n=1}^{N}\sum_{j=1}^2 \varepsilon_{j}c_{jn}^{\dagger}c_{jn}+\sum_{n=1}^{N-1}\sum_{j=1}^2 t_{j}c_{jn}^{\dagger}c_{jn+1}\nonumber\\
&+&\sum_{n=1}^{N-1}\sum_{j=1}^2 it_{\mathrm{SO}}c_{jn}^{\dagger}[\sigma_{n}^j+\sigma_{n+1}^j]c_{jn+1}\nonumber\\
&+&\sum_{n=1}^{N}\lambda c_{1n}^{\dagger}c_{2n}+H.c.
\label{eq:1}
\end{eqnarray}
with $c_{jn}^\pm=(c_{jn\uparrow}^\pm, c_{jn\downarrow}^\pm)$ denoting the spinor creation or annihilation operator on strand j at site n. The helical SOC is encoded by $\sigma_{n}^j=\sigma_{\perp}(\varphi_{n-1}+(j-1)\pi)$, where $\sigma_{\perp}(\varphi)=\sigma_xsin\varphi sin\theta-\sigma_ycos\varphi sin\theta+\sigma_zcos\theta$, with $\varphi_n=n\Delta\varphi$\cite{guo_spin-selective_2012,eremko_spin_2013}. Note that the SOC appears as a correction to the transfer integral in Eq.~\ref{eq:1}, consistent with the time-reversal symmetry (TRS) constraint that forbids on-site SOC terms in the absence of external magnetic fields\cite{sun_quantum_2005,guo_spin-selective_2012,entin-wohlman_comment_2021}. The donor–molecule coupling is $H_{DM}=\varepsilon_{D}c_{D}^{\dagger}c_{D}+\sum_{j}t_{DM}c_{D}^{\dagger}c_{1j}+H.c.$, connecting only to the first sites of the helix. The acceptor is described by $H_{MA}=\varepsilon_{A}c_{A}^{\dagger}c_{A}$, with its coupling to the helix included via the Lindblad jump operator defined below. 

The dynamics of the reduced density matrix $\rho(t)$ is governed by the Lindblad master equation ($\hbar=1$):
\begin{eqnarray}
\hbar\frac{d\rho}{dt}=-i[H_S,\rho]+\Gamma(L_A\rho L_A^{\dagger}-\frac{1}{2}\{L_A^{\dagger}L_A,\rho\})\nonumber\\
+\Gamma_H\sum_{n=1}^N\sum_{j=1}^2(L_{jn}\rho L_{jn}^{\dagger}-\frac{1}{2}\{L_{jn}^{\dagger}L_{jn},\rho\}).
\label{eq:2}
\end{eqnarray}
The removal of electrons from the molecular bridge to the acceptor is modeled by the jump operator $L_A=c_{A}^{\dagger}(c_{1N}+c_{2N})$, characterized by the rate $\Gamma$\cite{zhang_dynamical_2025,wojtowicz_open-system_2020}. In addition, pure dephasing arising from electron–phonon scattering, local defects, or other environmental sources is introduced via the operator $L_{jn}=c_{jn}^{\dagger}c_{jn}$, with strength $\Gamma_H$\cite{golizadeh-mojarad_nonequilibrium_2007,datta_electronic_1995,matityahu_spin-dependent_2016,fransson_vibrational_2020,fransson_charge_2021}. The donor is initialized in an equal superposition of spin-up and spin-down states, $\rho(0)=(|\uparrow\rangle\langle\uparrow|+|\downarrow\rangle\langle\downarrow|)_D/2$, unless otherwise state.

\begin{figure}[t]
\includegraphics[width = 1.0\columnwidth]{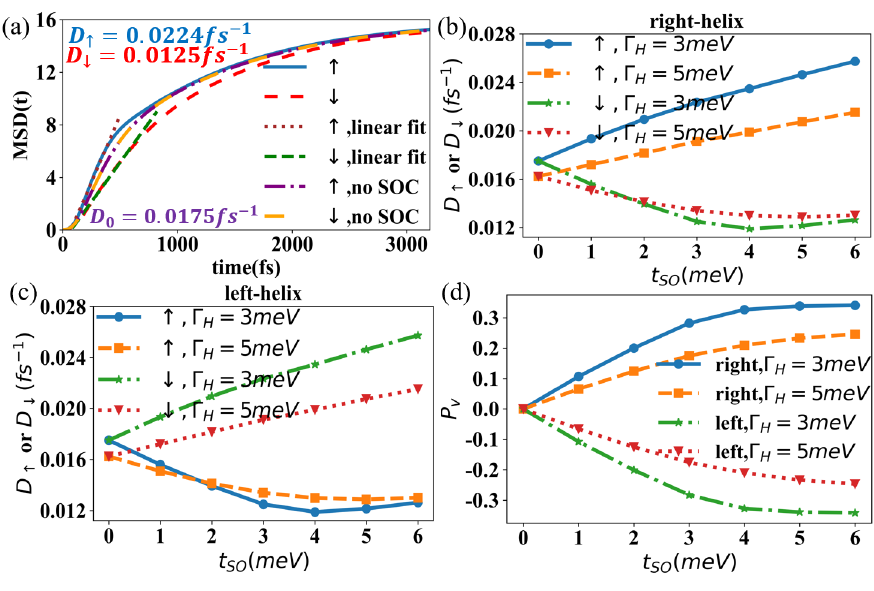}
\caption{\label{fig:2} (a) Spin-resolved MSD dynamics with and without SOC. Linear fits and the extracted slopes are indicated for both spin channels and the SOC-free baseline. (b) Spin‑resolved diffusion coefficients $D_{\uparrow}$ and $D_{\downarrow}$ as functions of $t_{\mathrm{SO}}$ for right‑handed helix. The plot reveals the “one-enhanced, one-suppressed” behavior and the non-monotonic trend of the suppressed spin branch. (c) For the left-handed helix, the two spin channels are reversed. (d) $P_v$ is plotted against $t_{\mathrm{SO}}$ for both helical configurations, with opposite signs and sizable values.}
\end{figure}

To quantify spin-selective transport, we define the spin-resolved mean-square displacement (MSD). In the diffusive regime where MSD grows linearly in time [Fig.~\ref{fig:2}(a)], the spin-resolved diffusion coefficient and mobility polarization are extracted from the MSD as
\begin{eqnarray}\label{eq:3}
\mathrm{MSD}_s(t)&=&\frac{1}{N}\sum_{j=1}^2\sum_{n=0}^{N+1} {{p_{jns}}(t){n^2}}\nonumber\\
D_s&=&\frac{d}{dt}\mathrm{MSD}_s(t)|_{linear}\nonumber\\
P_v&=&\frac{D_{\uparrow}-D_{\downarrow}}{D_{\uparrow}+D_{\downarrow}}
\end{eqnarray}
Here, $p_{jns}(t)=Tr(\rho(t) c_{jns}^{\dagger}c_{jns})$ denotes the occupation probability on strand j at site n for spin s. This definition follows the same form as that used in the analysis of single-stranded systems\cite{gelin_efficient_2021}. The linear regime used to extract $D_s$ is well-defined in diffusive transport\cite{wang_Communications_2010} and is illustrated in Fig.~\ref{fig:2}(a).

Throughout this work, we use parameters (in eV) adapted from Ref.~\cite{guo_spin-selective_2012,zhang_dynamical_2025}: $\varepsilon_1=0,\varepsilon_2=0.3,t_1=0.12,t_2=-0.1,\lambda=-0.3,t_{\mathrm{SO}}=0.003,\varepsilon_D=0.468,\varepsilon_A=-0.1,t_{DM}=0.04$ with structural parameters $\theta=0.66, \Delta\varphi=\pi/5$ and chain length $N=30$, unless otherwise stated. The dissipation rates $\Gamma$ and $\Gamma_H$ are chosen on the turnover plateau of the total diffusion coefficient (see Supplemental Material, Fig.S1\cite{SM}) with $\Gamma=0.3\mathrm{eV}$ and $\Gamma_H=3\mathrm{meV}$ unless otherwise stated, ensuring that the intrinsic transport properties of the molecule govern the dynamics\cite{wojtowicz_open-system_2020,Gruss_Communication_2017}. 

\emph{The One-Enhanced, One-Suppressed Phenomenon} --- We now present the "one-enhanced, one-suppressed" phenomenon previewed in the Introduction part. Fig.~\ref{fig:2}(a) shows the spin-resolved MSD dynamics for the right-handed helix with and without SOC. The MSD displays four regimes: departure from donor, wave-packet propagation along the bridge (the linear regime defining $D_s$), deviation upon reaching acceptor, and saturation after full transfer. In the presence of SOC, the two spin channels bifurcate. One is enhanced and the other is suppressed relative to the SOC-free baseline, yielding a steady-state mobility polarization of up to 28.3\% at $t_{\mathrm{SO}}=3meV$.

This phenomenon defies the naive expectation that such weak SOC, given its transfer-integral-like nature in Eq.~\ref{eq:1}, would merely produce a tiny enhancement of both spin channels. Instead, the SOC actively hinders one spin in absolute terms while promoting the other. Quantitatively, relative to the SOC-free case, $D_{\uparrow}$ increases by about \textbf{28\%} while $D_{\downarrow}$ decreases by the same amount, a nearly symmetric bifurcation phenomenon.

Fig.~\ref{fig:2}(b) and (c) display the extracted spin-resolved diffusion coefficients $D_{\uparrow}$ and $D_{\downarrow}$ as functions of $t_{\mathrm{SO}}$. For the right-handed helix [Fig.~\ref{fig:2}(b)], $D_{\uparrow}$ increases with $t_{\mathrm{SO}}$, while $D_{\downarrow}$ initially decreases, a genuine "one-enhanced, one-suppressed" phenomenon. The suppressed spin is non-monotonic. It reaches a minimum at intermediate $t_{\mathrm{SO}}$ and then recovers at larger values. This non-monotonic behavior reflects the dual role of SOC, which suppresses initially but then partially restores the mobility as the SOC amplitude grows. For the left-handed helix [Fig.~\ref{fig:2}(c)], the two spin channels are exchanged.

The resulting spin mobility polarization $P_v$ [Fig.~\ref{fig:2}(d)] reaches substantial values with opposite signs for the two helical structures. Even at $\mathbf{t_{\mathrm{SO}}=1meV}$, two orders of magnitude smaller than the hopping integrals, the polarization still reaches over \textbf{10\%}.

\emph{Microscopic Origin of CISS} --- We now trace the analytical origin of the "one-enhanced, one-suppressed" bifurcation phenomenon, which is also the microscopic origin of CISS. 

\begin{figure}[b]
\includegraphics[width = 1.0\columnwidth]{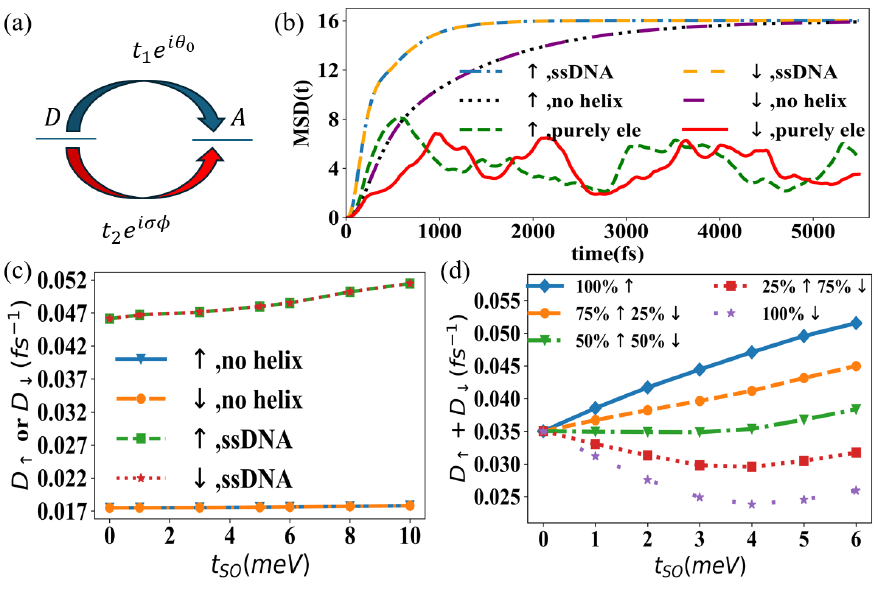}
\caption{\label{fig:3} (a) Schematic of the two-path interference model with spin-independent phase $\theta_0$ and SOC-induced phase $\phi$ adapted from \cite{sun_spontaneous_2005}. (b) MSD dynamics for three control cases: ssDNA and non-helical duplexes (no bifurcation), purely electronic model (oscillatory, no steady-state polarization). (c) Spin-resolved diffusion coefficients vs $t_{\mathrm{SO}}$ for ssDNA and non-helical duplexes, with only a tiny, symmetric enhancement of both spin channels. (d) Total diffusion coefficient $D_{\uparrow}+D_{\downarrow}$ vs $t_{\mathrm{SO}}$ as a function of the initial spin injection ratio.}
\end{figure}

Consider a minimal model where an electron traverses the helix via two paths [Fig.~\ref{fig:3}(a)]. In the absence of SOC, the total transmission is $T_0=|t_1e^{i\theta_0}+t_2|^2$ where $\theta_0$ is the spin-independent phase difference between the two paths. Without loss of generality, a tiny helical SOC adds a spin‑dependent phase $\sigma\phi$ to one of the path (e.g., path 2), without appreciably modifying its amplitude, while the other path remains unaffected\cite{sun_spontaneous_2005}. The SOC-induced phase $\phi$ is proportional to $t_{\mathrm{SO}}$\cite{sun_spontaneous_2005} and can accumulate along the helical paths (see Supplemental Material\cite{SM} for the full estimate). The spin-resolved transmission probabilities then become $T_{s}=|t_1e^{i\theta_0}+t_2e^{\pm i\phi}|^2$. Expanding this expression to second order in $\phi$ (valid for $|\phi|<0.4rad$) gives
\begin{eqnarray}\label{eq:4}
&T_{\uparrow}=T_0+2|t_1t_2|[\phi sin\theta_0-cos\theta_0(\frac{\phi^2}{2})]+O(\phi^3)\nonumber\\
&T_{\downarrow}=T_0+2|t_1t_2|[-\phi sin\theta_0-cos\theta_0(\frac{\phi^2}{2})]+O(\phi^3)
\end{eqnarray}

This is the microscopic origin of the "one-enhanced, one-suppressed" phenomenon. The linear terms carry opposite signs for the two spins. In the diffusive regime accessed by our MSD analysis, the spin-resolved diffusion coefficients are monotonically related to the transmission coefficients\cite{datta_electronic_1995,imry_conductance_1999}. The opposite signs in Eq.~\ref{eq:4} therefore directly translate to the "one-enhanced, one-suppressed" bifurcation of spin-resolved mobility observed in Fig.~\ref{fig:2}. The same expression also accounts for the recovery of the suppressed spin at larger $t_{\mathrm{SO}}$ (Fig.~\ref{fig:2}(b) and (c)). As $\phi$ grows with $t_{\mathrm{SO}}$, the increase of $|t_2|$ itself or higher-order terms begin to counteract the linear suppression, leading to the non-monotonic behavior.

From Eq.~\ref{eq:4}, the polarization follows directly. Assuming equal path amplitudes $|t_1|=|t_2|=|t|$, the polarization is then $P=\frac{T_{\uparrow}-T_{\downarrow}}{T_{\uparrow}+T_{\downarrow}}=\frac{sin\theta_0sin\phi}{1+cos\theta_0cos\phi}$. Crucially, the phase $\phi$ is of order $0.2rad$ (see Supplemental Material\cite{SM} for the estimate), \textbf{small as an angle, but by no means negligible when compared to unity}. The mobility polarization can be even amplified via the nonlinear mapping $D_s\propto \frac{T_s}{1-T_s}$\cite{imry_conductance_1999,datta_electronic_1995}. 

The control results in single-stranded DNA or non-helical duplexes [Fig.~\ref{fig:3}(b) and (c)] further confirm our theoretical picture. In those cases, the same weak SOC then yields no spin-resolved mobility difference. At $t_{\mathrm{SO}}=3meV$, the same SOC strength used in the dsDNA, both spin diffusion coefficients increase by less than \textbf{2.5\%}, in stark contrast to the \textbf{28\% suppression} and enhancement in the dsDNA case. Moreover, SOC must be increased from 3meV to about 10 meV before any detectable MSD change appears, consistent with the naive expectation that weak SOC merely produces a tiny mobility enhancement. This is consistent with experimental observations that single-stranded DNA shows no CISS effect\cite{gohler_spin_2011,guo_spin-selective_2012}. 

The purely electronic model in Fig.~\ref{fig:3}(b) serves as an additional control. Without dissipation, the asymmetry oscillates and averages to zero, confirming that dissipation is essential for locking the transient asymmetry into a steady-state polarization. Together, these controls establish that helical SOC, multi-path interference, and dissipation are all essential for the emergence of steady-state spin polarization.

The "one-enhanced, one-suppressed" asymmetry uncovered above can also provide a natural explanation for the observation that the total mobility can be tuned by spin injection. From Eq.~\ref{eq:4}, the linear SOC corrections to the individual spin transmission cancel in the sum $T_{\uparrow}+T_{\downarrow}$, leaving the total transmission only weakly dependent on SOC. The total mobility can in turn be tuned by the initial spin injection without affecting the polarization. As shown in Fig.~\ref{fig:3}(d), the total diffusion coefficient $D_{\uparrow}+D_{\downarrow}$ for right-handed dsDNA is highly sensitive to the initial spin composition. A larger fraction of the injected favored spin increases the overall mobility, whereas a larger fraction of the disfavored spin reduces it. Mixed spin injections yield intermediate mobility between these two limits. This is a direct signature of mobility polarization and a pure spin population asymmetry would not change the total current. Existing experiments measuring the electronic current with ferromagnetic contacts have already observed such a dependence\cite{naaman_chiral_2019,naaman_chiral_2020,SM}, consistent with our discovery. Thus, when properly initialized, the donor–acceptor system can achieve both high charge mobility and strong spin selectivity.

\begin{figure}[t]
\includegraphics[width = 1.0\columnwidth]{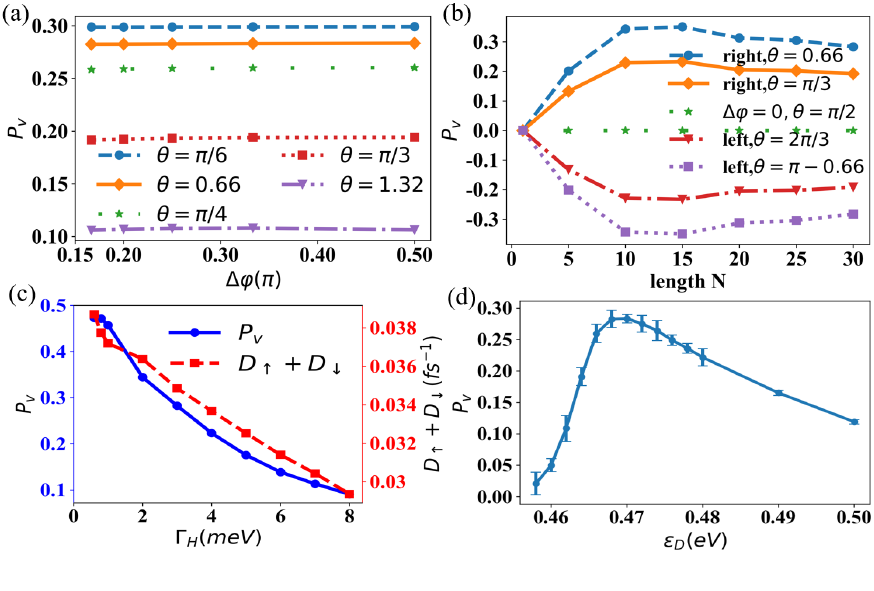}
\caption{\label{fig:4} (a) Mobility polarization $P_v$ plotted against the helical parameters $\theta$ and $\Delta\varphi$. (b) Dependence of $P_v$ on chain length N, comparing achiral, right-handed, and left-handed configurations. (c) $P_v$ and $D_{\uparrow}+D_{\downarrow}$ vs $\Gamma_H$. (d) $P_v$ vs donor energy $\epsilon_D$, showing appreciable values over a broad energy range.}
\end{figure}

\emph{Structural and Parameter Dependence} --- We now examine the dependence of the mobility polarization on geometry, size, environmental dissipation, and donor energy. Fig.~\ref{fig:4}(a) shows $P_v$ as a function of the twist angle $\Delta\varphi$ for several helix angles $\theta$ in the right-handed helix configuration.  Increasing $\theta$ leads to a monotonic decrease in the polarization. In contrast, varying $\Delta\varphi$ over a wide range, from $\pi/6$ to $\pi/2$, has only a marginal influence on $P_v$, suggesting that the helix angle $\theta$ serves as the dominant geometric parameter shaping the effective SOC field.

Fig.~\ref{fig:4}(b) shows $P_v$ versus the chain length N for right-handed, left-handed, and achiral structures. The polarization grows with N and and levels off for  $N\geq15$, , consistent with the length-dependent spin selectivity observed in earlier experiments\cite{guo_spin-selective_2012,gohler_spin_2011}. Achiral dsDNA shows no polarization at any length\cite{gohler_spin_2011,guo_spin-selective_2012}, further validating our theoretical picture.

We now turn to the effect of environmental dissipation. Fig.~\ref{fig:4} (c) shows $P_v$ as a function of $\Gamma_H$. The polarization decreases with increasing $\Gamma_H$, but remains appreciable over a broad range, indicating that the "one-enhanced, one-suppressed" phenomenon is robust against variations in dissipation. The complete dependence of $P_v$ on $\Gamma$ and $\Gamma_H$ is provided in the Supplemental Material\cite{SM}.

Fig.~\ref{fig:4} (d) shows $P_v$ as a function of the donor energy $\epsilon_D$. The polarization is very small when $\epsilon_D$ is off resonance, rises sharply as the donor level enters the resonance window (from diagonalization of the isolated bridge Hamiltonian), which also enhances the overall charge transfer efficiency, and reaches a maximum around 0.468–0.470 eV. Above the peak, $P_v$ varies slowly over a wide range. Within a moderate window of donor energy, $P_v$ changes by only a few percent, indicating that the polarization is robust against typical uncertainties in donor energy.

\emph{Conclusions} --- We have uncovered a counterintuitive phenomenon, a “one-enhanced, one-suppressed” bifurcation of spin-resolved mobility. It also constitutes the microscopic origin of CISS. We trace this bifurcation analytically to a sign reversal, resolving the core puzzle of whether, how and why weak SOC within a helical molecule produces large spin polarization.

We further show that the bifurcation phenomenon requires helicity, SOC, multi-path interference, and dissipation to act together by control results. Without any one of them, the steady bifurcation does not occur, and SOC only produces a weak, symmetric enhancement, consistent with the naive expectation. This theoretical picture can also show that the total mobility can be tuned by spin injection without affecting spin selectivity, a signature that is already consistent with existing experiments and directly testable. The origin we have uncovered is not limited to dsDNA but applies to any helical chiral systems.

\begin{acknowledgments}
\emph{Acknowledgments} --- This work is supported by the National Natural Science Foundation of China (Grant Nos. T2350009 and 22433007), the Guangdong Provincial Natural Science Foundation (Grant No. 2024A1515011185), and the Shenzhen City Peacock Team Project (Grant No. KQTD20240729102028011). Z. Shuai and W. Li are supported by the Guangdong Basic Research Center of Excellence for Aggregate Science.
\end{acknowledgments}




\bibliography{apssamp}

\onecolumngrid

\clearpage

\makeatletter
\renewcommand{\c@secnumdepth}{0}
\makeatother
\renewcommand{\thesection}{\Alph{section}}

\renewcommand{\thefigure}{\Alph{section}.\arabic{figure}}
\renewcommand{\thetable}{\Alph{section}.\arabic{table}}
\renewcommand{\theequation}{\Alph{section}.\arabic{equation}}
\setcounter{figure}{0}
\begin{center}
  {\bf\large Supplemental Material for}\\[1ex]
  {\bf\large “The Counterintuitive ‘One-Enhanced, One-Suppressed’ Bifurcation Phenomenon of Spin Mobility under Weak Spin-Orbit Coupling in Helice"}\\
\end{center}
\setcounter{equation}{0}
\setcounter{figure}{0}
\section{Why Weak SOC Induces Large Polarization: An Estimate of the SOC-Induced Phase}\label{ssec:A}
In the main text, the SOC-induced phase $\phi$ is described as “modest as an angle, but by no means negligible when compared to unity.” This is the key to understanding why a weak SOC, with $t_{\mathrm{SO}}$ about two orders of magnitude smaller than the hopping integrals, can still induce a large spin polarization. Here we provide the quantitative estimate behind this statement. 

Following Ref.~\cite{sun_spontaneous_2005}, the phase $\phi$ can be estimated as $\phi=k_RL$, with $k_R=m^*\alpha/\hbar^2$. $\alpha=4*t_{\mathrm{SO}}*l_a$ is is the effective SOC coefficient, $l_a\approx0.56\mathrm{nm}$ is the arc length between two base pairs in one chain, and L is the effective path length. Taking $L=2N*l_a$ (L can be even larger since there are more paths, and the length between two chains is longer than $2l_a$\cite{guo_spin-selective_2012}), and taking $m^*=0.03-0.1 m_e$, we obtain $\phi=0.0887-0.296$. This range falls within the range used in the error analysis, ensuring that the second-order expansion in Eq. 4 of the main text is valid. This is why weak SOC-induced spin polarization is not tiny, with $\phi$ small as an angle, but not negligible when compared to unity.

Substituting $\phi\approx0.3rad$ into the estimate $P=\frac{T_{\uparrow}-T_{\downarrow}}{T_{\uparrow}+T_{\downarrow}}=\frac{2|t_1t_2|sin\theta_0sin\phi}{|t_1|^2+|t_2|^2+2|t_1t_2|cos\theta_0cos\phi}$ and assuming equal path amplitudes $|t_1|=|t_2|$, gives $P=17.32\%$ for $\theta_0=\pi/3$ and $P=30\%$ for $\theta_0=\pi/2$, consistent with the numerical polarization values in the main text. Actually, since the effect of SOC on $|t_2|$ is tiny, the equal-amplitude case $|t_1|\approx|t_2|$ can be seen as a natural starting point. Even if the amplitudes are not exactly equal, the qualitative picture remains unchanged.

We also note an additional amplification mechanism. In the diffusive regime, the spin-resolved diffusion coefficients are related to the transmission probability by the nonlinear mapping $D_s\propto \frac{T_s}{1-T_s}$\cite{datta_electronic_1995,imry_conductance_1999}. This mapping amplifies small differences in transmission into larger differences in mobility.

The expression for P also reveals a design principle. The polarization can be further enhanced if $cos\theta_0<0$. In this case, since $T_{\uparrow}+T_{\downarrow}=2T_0+2|t_1t_2|cos\theta_0(cos\phi-1)$, it enhances the total transmission as well. Since $\theta_0$ is determined by the molecular geometry (e.g., helix angle and inter-chain coupling),  the condition $cos\theta_0<0$ is achievable by structural design, offering a way to enhance both spin polarization and charge mobility in chiral systems.

\setcounter{equation}{0}
\setcounter{figure}{0}
\section{Dissipation Parameter Selection: The Turnover Plateau}\label{ssec:B}
Here we describe the parameter selection of the dissipation parameters in the main text. As described in the General Setup section, we include two dissipation channels: the irreversible electron transfer to the acceptor with rate $\Gamma$, and local pure dephasing with strength $\Gamma_H$. Both dissipation channels are spin-independent in Eq. 2.
\begin{figure}[ht]
\includegraphics[width = 0.7\columnwidth]{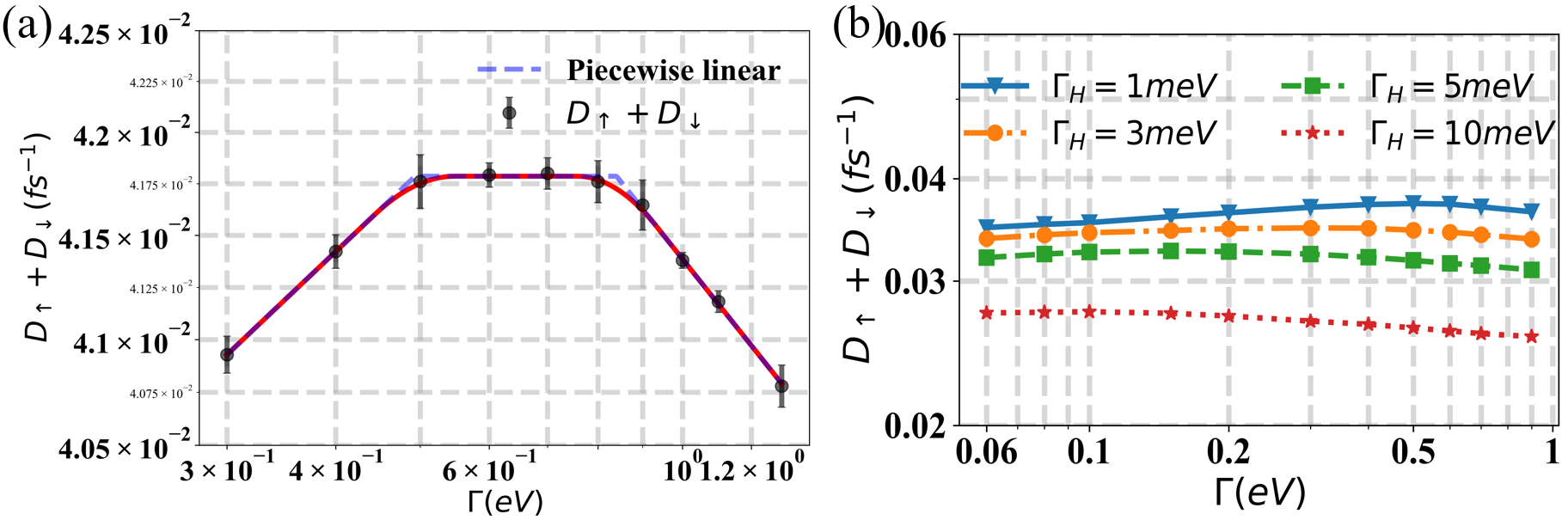}
\caption{\label{fig:b1} (a) Total diffusion coefficient vs $\Gamma$, showing the turnover plateau. (b) Total diffusion coefficient vs $\Gamma$ for different $\Gamma_H$, illustrating many-body effects on the turnover induced by dissipation $\Gamma_H$.}
\end{figure}

Fig.~\ref{fig:b1}(a) shows the total diffusion coefficient $D_{\uparrow}+D_{\downarrow}$ as a function of $\Gamma$ with $\Gamma_H=0$. The curve exhibits a turnover behavior: a rise at small $\Gamma$, a plateau at intermediate $\Gamma$, and a decay at large $\Gamma$. The behavior is analogous to the Kramers turnover in chemical reaction rates\cite{wojtowicz_open-system_2020,Gruss_Communication_2017}. At small $\Gamma$, the transfer out of the molecule is slow, limiting the overall charge flux. At large $\Gamma$, the extremely fast removal of electrons suppresses the quantum coherence that is necessary for establishing a steady‑state mobility. Only on the plateau does the intrinsic molecular dynamics\cite{wojtowicz_open-system_2020} govern transport. We therefore choose $\Gamma=0.6 eV$ inside the plateau region for $\Gamma_H=0$, which ensures the steady‑state mobility polarization is well defined.

The relaxation rate $\Gamma_H$ can be introduced as a phenomenological parameter representing the effect of spin‑independent electron‑phonon scattering processes. Fig.~\ref{fig:b1}(b) shows $D_{\uparrow}+D_{\downarrow}$ as a function of $\Gamma$ for several values of $\Gamma_H$. The same turnover behavior is observed, with the entire turnover curve shifting to lower $\Gamma$ as $\Gamma_H$ increases. The essential temperature dependence of $\Gamma_H$ can be motivated by phenomenological system‑bath coupling models for pure dephasing and takes the form\cite{guo_quantum_1990,Brasil_A_2013}
\begin{eqnarray}\label{eq:S1}
\Gamma_H(T)\propto J(w_0)coth(\frac{w_0}{2k_BT})
\end{eqnarray}
where $J(w)$ is the spectral density and $w_0$ is a characteristic phonon frequency. At high temperatures ($k_BT\gg \hbar w_0$), this reduces to $\Gamma_H \propto T$, while at low temperatures it approaches a constant due to zero-point fluctuations. This provides a physically reasonable basis for the temperature dependence of $\Gamma_H$ in our phenomenological Lindblad model. The specific value of $\Gamma_H=3\mathrm{meV}$ used in the main text is chosen phenomenologically. Its magnitude is consistent with the energy scale of low-frequency phonon modes in DNA, which have been reported in the range from sub-meV up to about 10 meV\cite{Blinov_acoustic_2010,woolard_submillimeter_2002}. Therefore, a value of $3meV$ is proper for $\Gamma_H$ at room temperature.

Fig.~\ref{fig:b2} shows the images of MSD curves for different $\Gamma$ and $\Gamma_H$ regimes. When $\Gamma$ and $\Gamma_H$ are tiny, the MSD curves shows clear electronic character, which are oscillatory and time-averaged-zero as shown below. Based on the behavior of the spin‑resolved MSD and the total diffusion coefficient, we identify the following regimes (without $\Gamma_H$, adding $\Gamma_H$ effectively increasing $\Gamma$):
\begin{figure}[ht]
\includegraphics[width = 1\columnwidth]{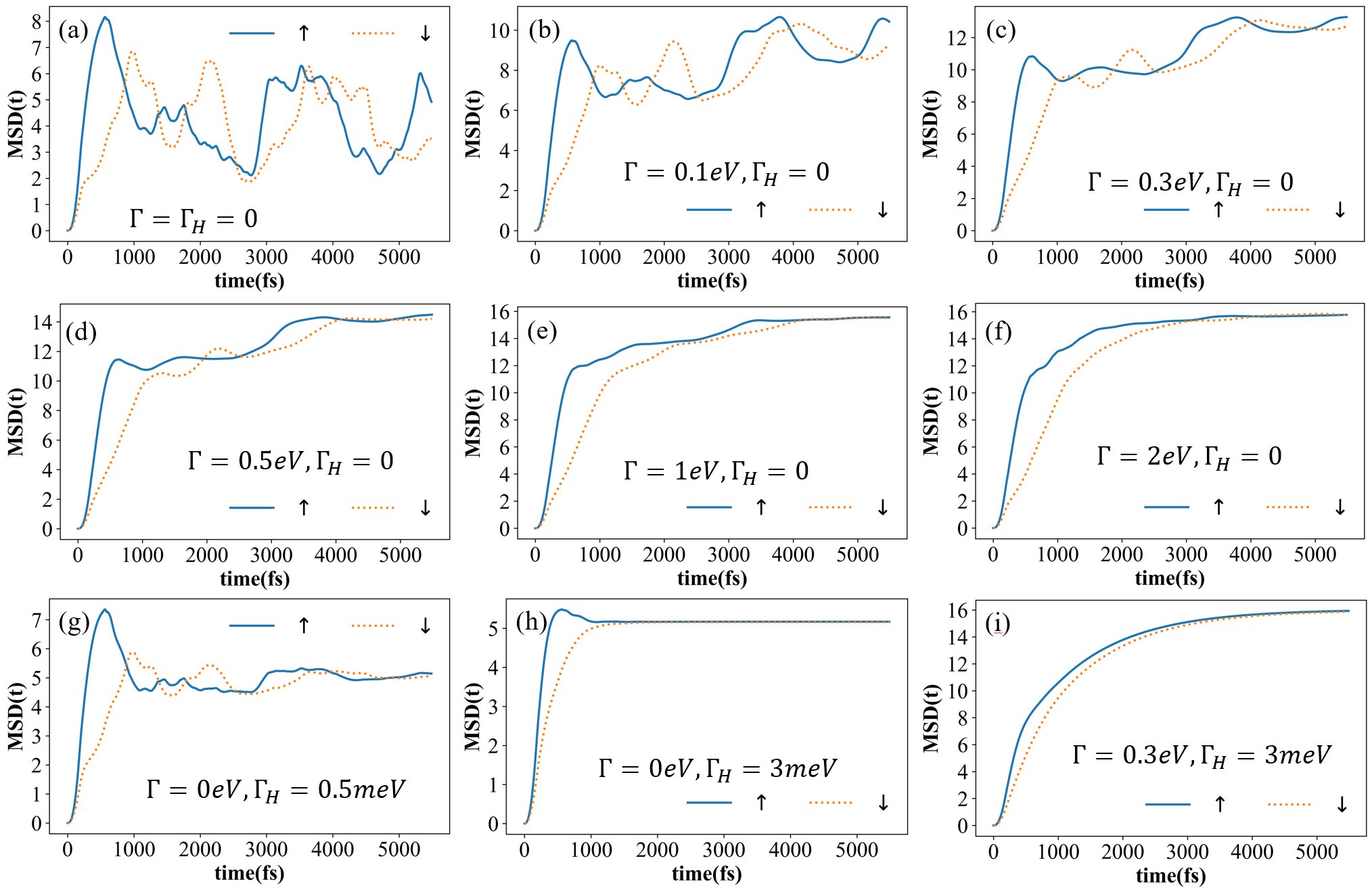}
\caption{\label{fig:b2} (a),(b),(c),(d),(e),(f) The MSD curves for different values of $\Gamma$ with $\Gamma_H=0$. (g),(h),(i) The MSD curves for different values of $\Gamma_H$ with $\Gamma$.}
\end{figure}
\begin{itemize}
\item{$\Gamma<0.1eV$:}
No steady state appears. The wave packet reflects multiple times, MSD lacks a linear regime. $D_{\uparrow}+D_{\downarrow}$ can not be reliably defined. $P_v$ is ill‑defined because the two spin‑resolved MSD curves alternately overtake each other. Consequently, the time‑averaged polarization tends to zero.
\item{$0.1eV\leq\Gamma<0.3 eV$:}
Steady state begins to emerge but oscillations remain strong. Extracted $D_{\uparrow}+D_{\downarrow}$ has large uncertainty. $P_v$ is still ill‑defined and the time‑averaged polarization is small.
\item{$0.3eV\leq \Gamma\leq0.5 eV$:}
Steady state is established. Total diffusion coefficient increases rapidly with $\Gamma$. Meanwhile, 
$P_v$ rises, reaches a maximum near $\Gamma\approx0.5eV$ as shown in Fig.~\ref{fig:c1}, and then starts to decline.
\item{$0.5eV< \Gamma\leq0.8 eV$:}
Plateau region appears. Diffusion coefficient is nearly constant, and mobility polarization can be reliably defined. $P_v$ decreases slowly after its peak.
\item{$\Gamma>0.8 eV$:}
This is the over‑damped regime. Diffusion coefficient decays, but steady state is still definable. $P_v$ continues to decrease.
\end{itemize}

\setcounter{equation}{0}
\setcounter{figure}{0}
\section{Polarization Dependence on Dissipation Parameters}\label{ssec:C}
\begin{figure}[ht]
\includegraphics[width = 0.4\columnwidth]{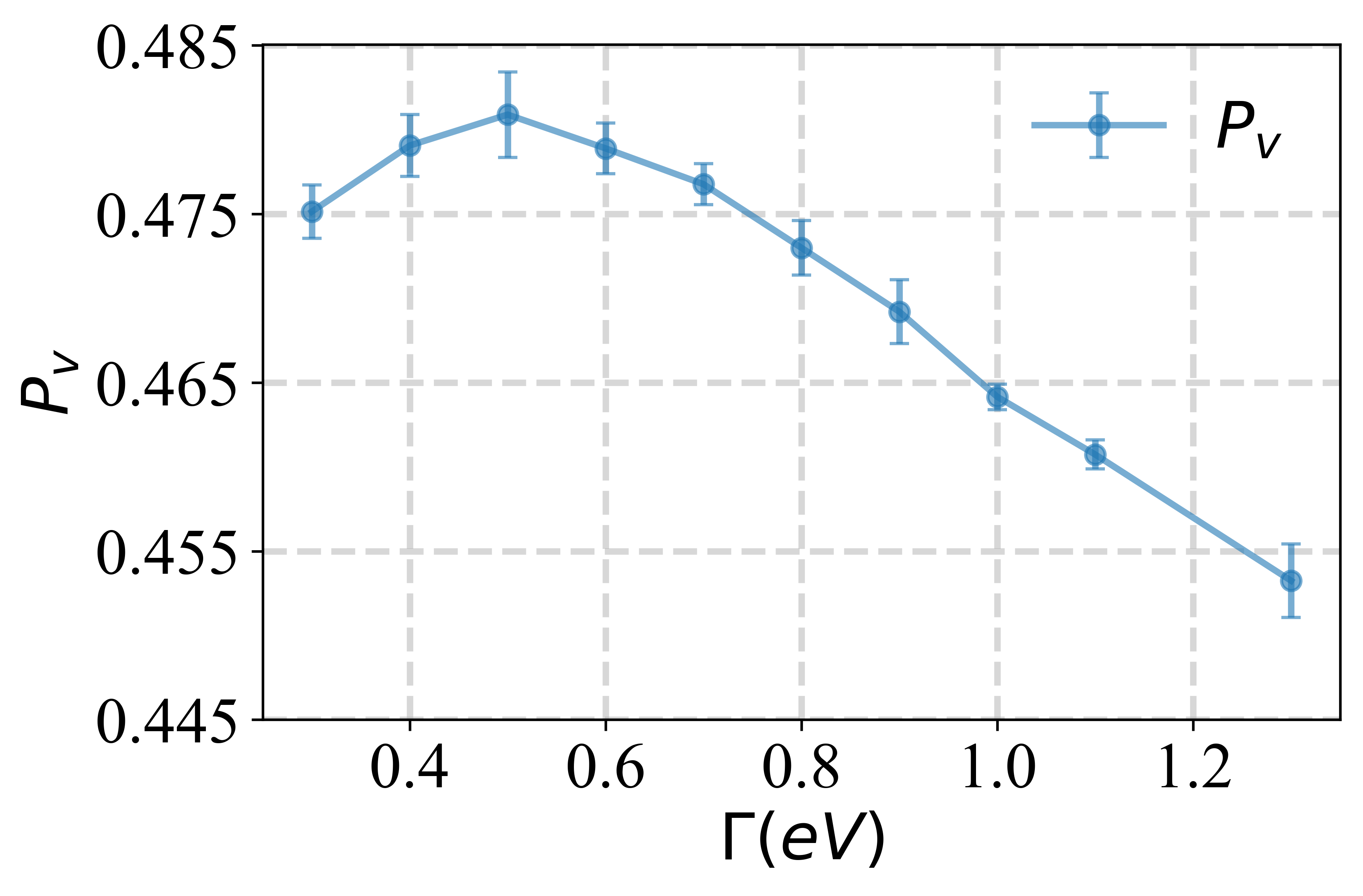}%
\caption{\label{fig:c1} Mobility polarization $P_v$ vs $\Gamma$ without $\Gamma_H$.}
\end{figure}
As shown in Fig. 4(c) in the main text, we show that the mobility polarization remains substantial over a broad range of dissipation parameters. The underlying qualitative mechanism is universal and independent of the specific value of parameters. Here we provide the complete dependence of $P_v$ on $\Gamma$ and $\Gamma_H$.

Fig.~\ref{fig:c1} shows the mobility polarization $P_v$ as a function of the electron transfer rate $\Gamma$ for $\Gamma_H=0$. The polarization first increases, reaches a maximum near $\Gamma\approx0.5eV$ at the onset of the plateau of the total diffusion coefficient (see Fig.~\ref{fig:b1}(a)), and then slightly decreases. This indicates that the optimal dissipation for polarization is achieved at the very beginning of the plateau regime. When we further increase $\Gamma$ into the plateau or additionally turn on $\Gamma_H$, the combined dissipation pushes the system into the reduced polarization regime.
\begin{figure}[ht]
\includegraphics[width = 0.4\columnwidth]{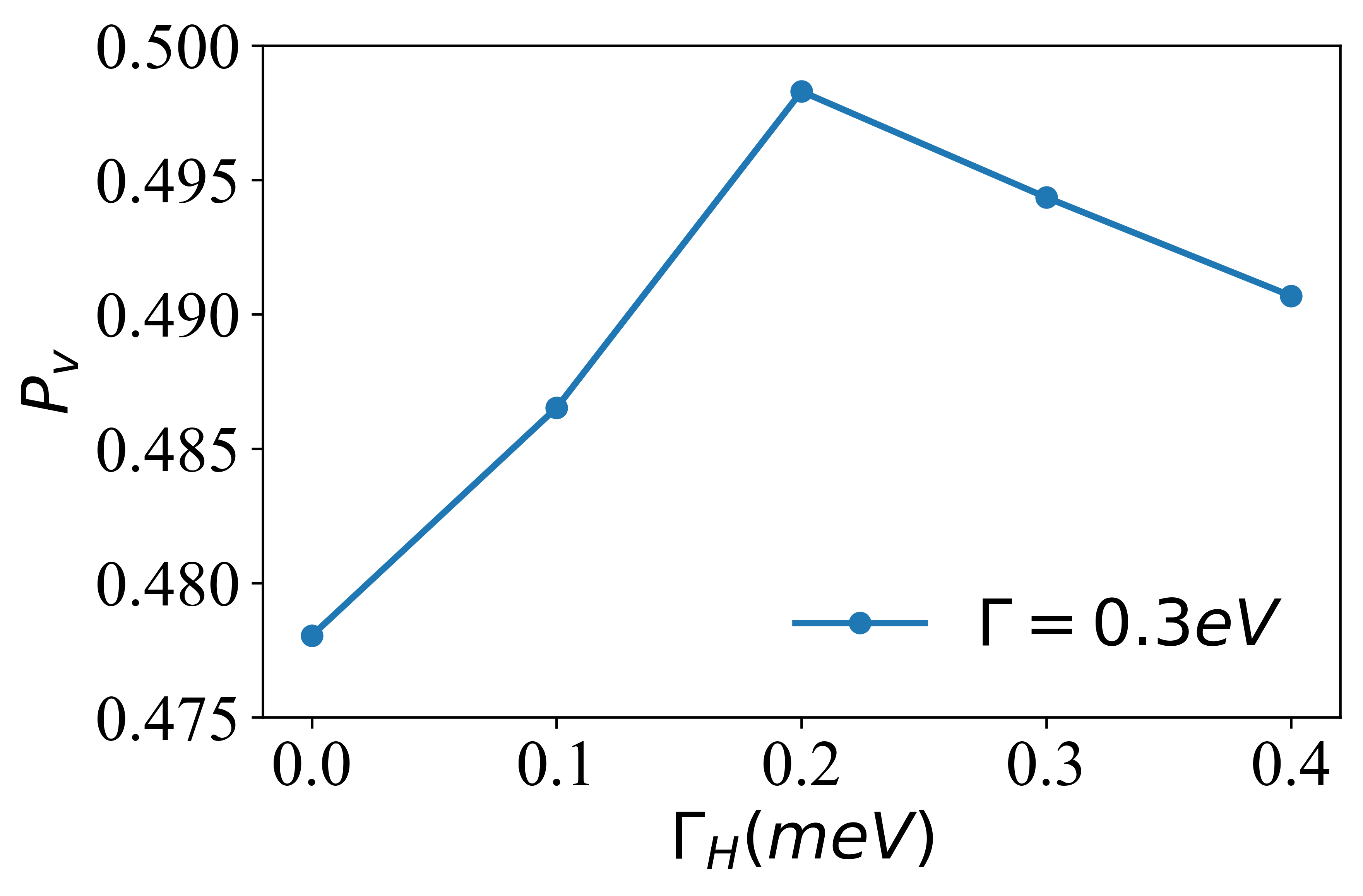}
\caption{\label{fig:c2} Mobility polarization $P_v$ vs $\Gamma_H$ with fixed $\Gamma=0.3eV$.}
\end{figure}

For a fixed small $\Gamma$, (e.g., $\Gamma=0.3eV$, lying to the left of the plateau), $P_v$ as a function of $\Gamma_H$ exhibits a non‑monotonic behavior. It first increases, reaches a maximum, and then decreases in Fig.~\ref{fig:c2}. This reflects the fact that a small amount of pure dephasing can enhance the spin asymmetry. In contrast, for our working point on the plateau in the main text, the polarization already lies on the decreasing branch. Thus, as shown in Fig. 4(c), $P_v$ decreases monotonically with increasing $\Gamma_H$. The observed monotonic decrease of $P_v$ with $\Gamma_H$ is consistent with experimental reports that CISS polarization decreases with increasing temperature in certain systems\cite{qian_chiral_2022,das_temperature-dependent_2022} and consistent with analysis from Ref.~\cite{liu_dynamical_2025}, where stronge dephasing suppresses the spin asymmetry.

The mobility polarization $P_v$ is further refined and analyzed in Supplemental Material, Sec.~\ref{ssec:D} by considering the alternating-to-fixed order locking effect.  An exact treatment of phonon spectral density, including its quantitative effects on the polarization, is beyond the scope of the present work.

\setcounter{equation}{0}
\setcounter{figure}{0}
\section{Leading‑Time Fraction: Quantifying the Alternation‑to‑Locking Transition}\label{ssec:D}
To quantify the transition from the alternating (oscillatory) asymmetry in the purely electronic model to the fixed‑order asymmetry and steady polarization under dissipation, we define the leading‑time fraction $f$
\begin{eqnarray}\label{eq:d1}
f = \frac{1}{T} \int_{0}^{T} \left[ \Theta\bigl(\mathrm{MSD}_{\uparrow}(t)-\mathrm{MSD}_{\downarrow}(t)\bigr) - \Theta\bigl(\mathrm{MSD}_{\downarrow}(t)-\mathrm{MSD}_{\uparrow}(t)\bigr) \right] dt,
\end{eqnarray}
where $\Theta$ is the Heaviside step function and $T$ is a finite time window chosen at which the acceptor population reaches e.g. 65\% (depending on the well-defined linear regime established) of the total population, to ensure a finite window and avoid the $1/T\rightarrow0$ limit. The quantity $f$ measures the net fraction of time during which one spin consistently leads the other. The behavior of $f$ can be divided into three regimes based on the dissipation strength $\Gamma$ ($\Gamma_H=0$):
\begin{itemize}
\item{$\Gamma\leq0.2 eV$:}
For the purely electronic model with almost zero $\Gamma$, where no steady state exists and acceptor population never saturates within the simulation time, $f$ is evaluated over the full simulation duration. The two MSD curves cross periodically, giving $f\approx0$, indicating no persistent leading spin.
\item{$0.2eV<\Gamma\leq0.4 eV$:}
For intermediate dissipation, the oscillations are partially suppressed but not fully eliminated. The MSD curves still cross occasionally, so $|f|$ takes moderate values between 0 and 1 ($|f|\approx0.3-0.6$), reflecting the gradual transition from alternating to fixed‑order.
\item{$\Gamma>0.4eV$:}
As dissipation becomes moderate, the oscillations are fully suppressed and one spin maintains almost a persistent lead. Accordingly, $f$ increases towards $+1$ when up-spin is ahead (right helix dsDNA) or $-1$ when down-spin is ahead (left helix dsDNA). For our working parameters, we find $|f|\rightarrow1$, confirming that the alternating asymmetry has been effectively locked into a steady polarization.
\end{itemize}

\begin{figure}[ht]
\includegraphics[width = 0.4\columnwidth]{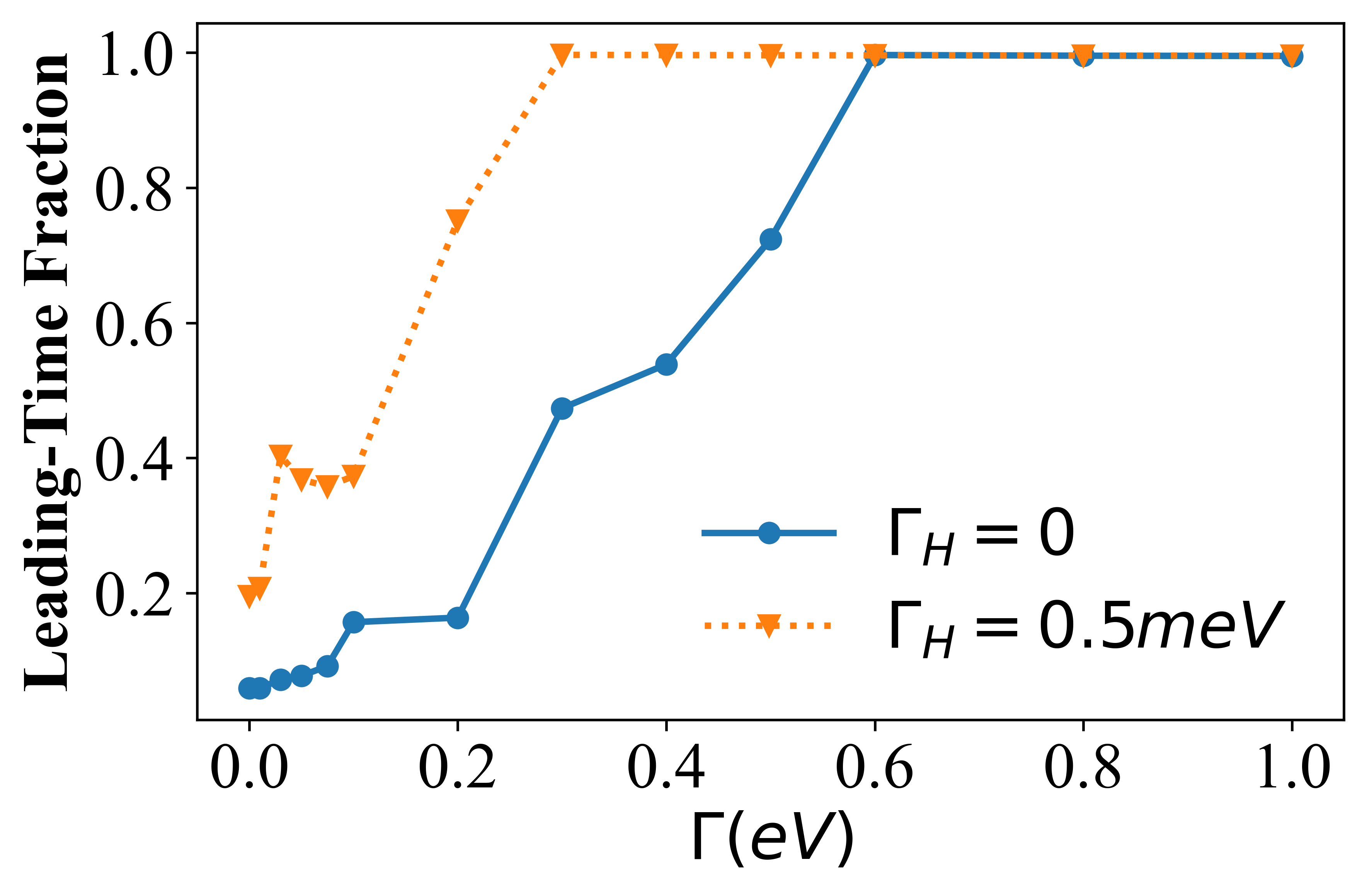}
\caption{\label{fig:d2} Leading‑time fraction $f$ as a function of $\Gamma$ with or without $\Gamma_H$ for right‑handed helix.}
\end{figure}

Fig.~\ref{fig:d2} shows $f$ vs $\Gamma$ for right‑handed helix. It rises from near zero (alternating) to unity (fixed locking) as $\Gamma$ increases. When $\Gamma_H$ is turned on, the same trend is observed but the $f$ curve shifts to smaller $\Gamma$.

The role of dissipation is most clearly revealed by comparing $|f|$ with and without it. If $f\approx0$ in the absence of dissipation, then a large $|f|$ upon adding dissipation indicates a genuine locking of the order. In contrast, if the purely electronic model already yields $|f|\approx1$ (one spin persistently leads), then the large $|f|$ after adding dissipation merely reflects smoothing of the curves, not a locking transition. Our results fall into the first category, identifying the locking effect of dissipation, instead of merely smoothing.

\begin{figure}[ht]
\includegraphics[width = 0.5\columnwidth]{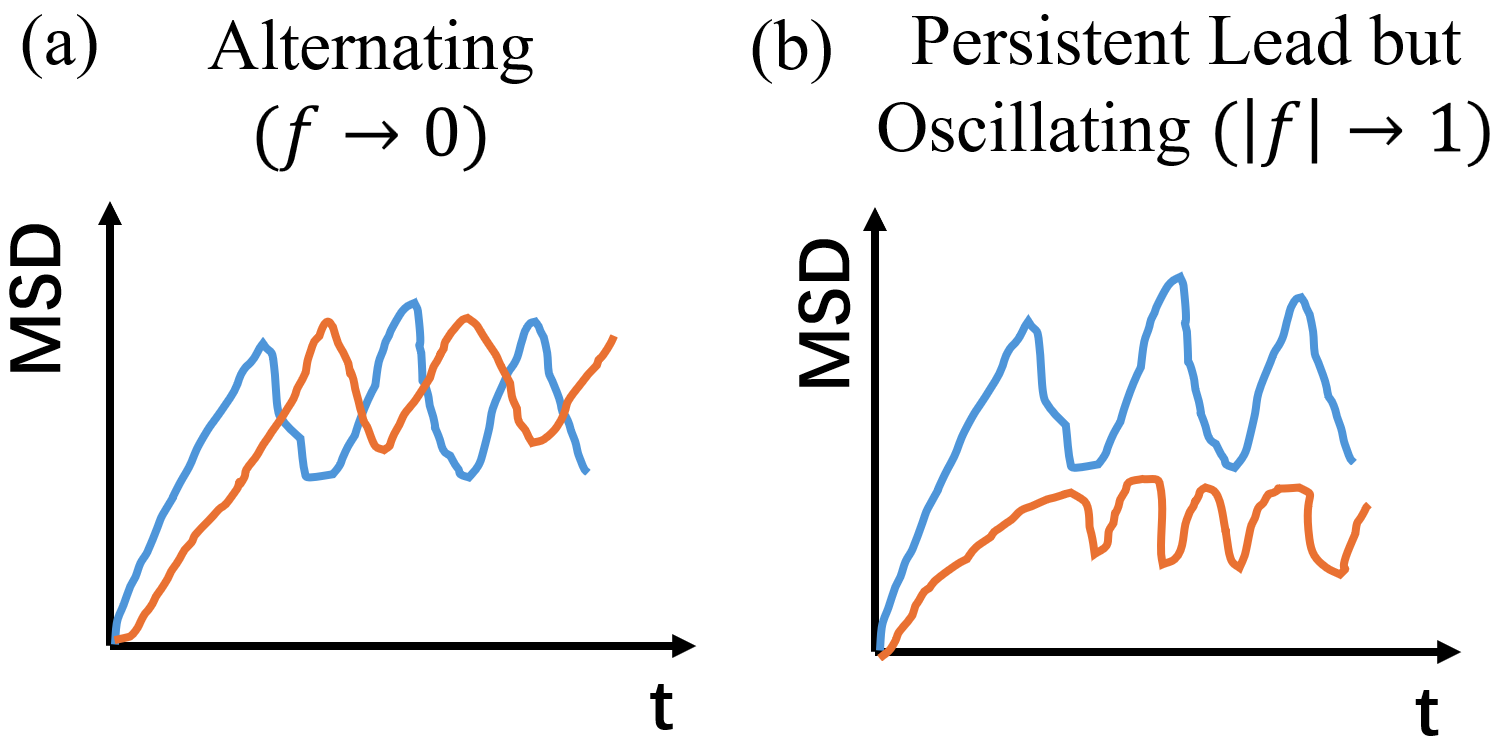}
\caption{\label{fig:d1} Schematic of two qualitatively distinct behaviors in the purely electronic model. (a) Alternating: the two MSD curves cross periodically ($f\approx0$). (b) Persistent lead: (e.g., TRS‑violating model), one spin always stays above the other, with oscillations but less crossing ($|f|\approx1$)}
\end{figure}

In principle, two qualitatively distinct behaviors, alternating crossing versus fixed order, can occur in the purely electronic model, as illustrated schematically in Fig.~\ref{fig:d1}. The purely electronic model yields $f\approx0$ or $|f|\approx1$. This distinction is not a mere theoretical construct. We have also numerically verified (results not included here) the latter scenario using a time-reversal symmetry (TRS)‑violating SOC (e.g., on‑site spin‑flip terms).
\begin{figure}[ht]
\includegraphics[width = 0.4\columnwidth]{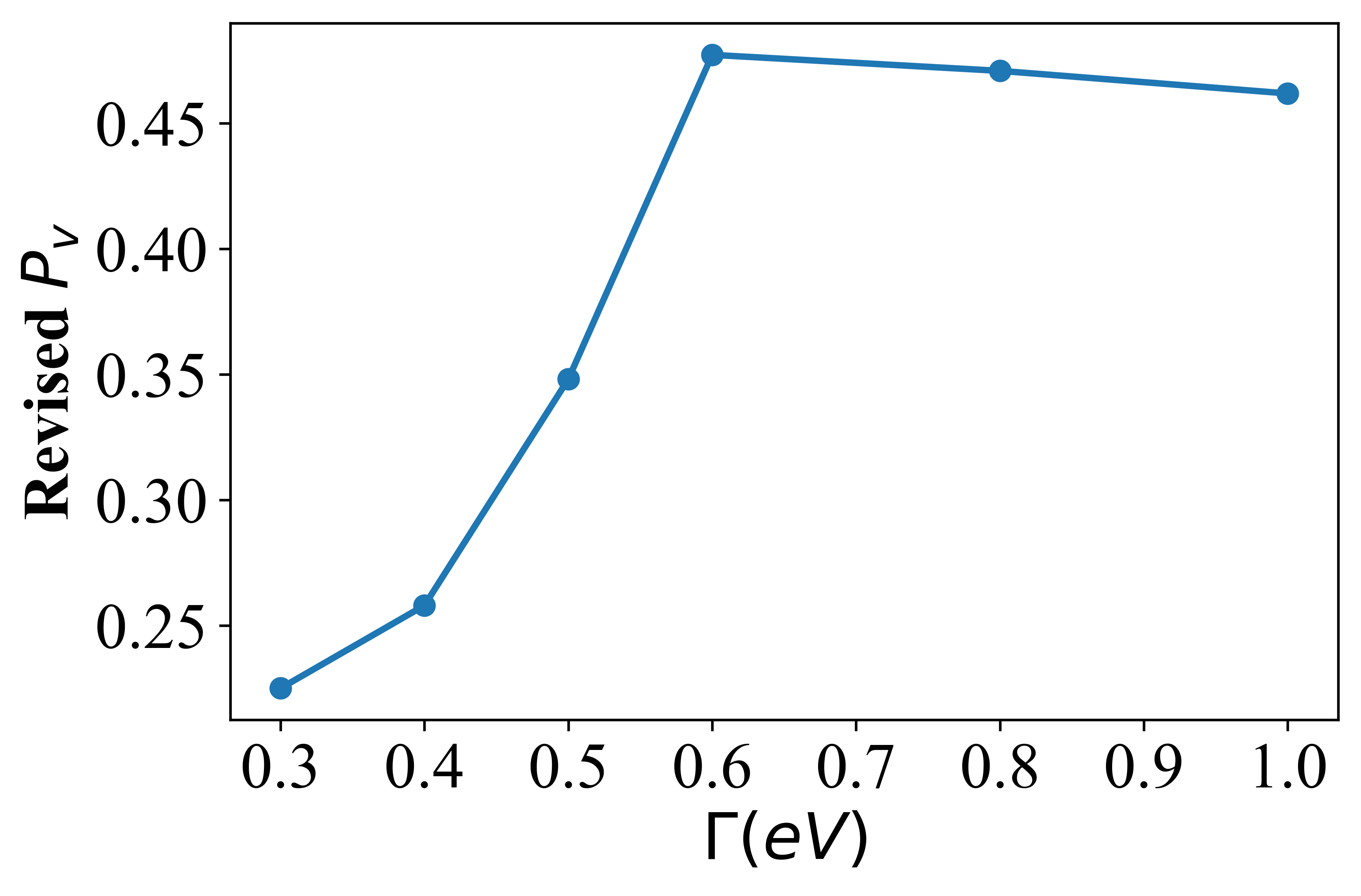}
\caption{\label{fig:d3} Revised spin mobility polarization $P_v^{\mathrm{revised}}=P_v * |f|$ as a function of $\Gamma(\Gamma_H=0)$ for the right‑handed helix.}
\end{figure}

Using the leading-time fraction, a revised mobility polarization can be defined as
\begin{eqnarray}\label{eq:d2}
P_v^{\mathrm{revised}}=P_v * |f|
\end{eqnarray}
where $P_v$ is the mobility polarization extracted from the MSD slope (as defined in the main text). This product accounts for the degree of alternating crossing or fixed‑order locking. In the fully locked regime ($|f|\rightarrow1$), it approaches original $P_v$. In the purely electronic or weakly dissipative regime ($|f|\rightarrow0$), it becomes small, reflecting that the transient difference does not produce a steady polarization. 

This simple metric provides a quantitative criterion for the transition from oscillation to dissipation‑locked spin selectivity, complementing the mobility polarization $P_v$ when a steady state cannot be reached and $P_v$ cannot be quantified. Fig.~\ref{fig:d3} shows the revised mobility polarization $P_v^{\mathrm{revised}}$ as a function of $\Gamma$. For small $\Gamma$, where $f\ll1$, the effective polarization is much reduced. As $\Gamma$ becomes larger and $|f|\rightarrow1$, $P_v^{\mathrm{revised}}$ approaches the original $P_v$. Thus, even spin‑independent dissipation plays an essential role. It converts the transient, incomplete transfer and oscillatory spin asymmetry of the purely electronic model (Fig.3(b)) into a steady‑state mobility polarization.

\setcounter{equation}{0}
\setcounter{figure}{0}
\section{Time‑Reversal Symmetry in Spin‑Orbit Coupling: Allowed and Forbidden Forms}\label{ssec:E}
The Hamiltonian must satisfy time‑reversal symmetry (TRS) in the absence of external magnetic fields. TRS imposes strict constraints on the matrix elements of the Hamiltonian\cite{sun_quantum_2005}:
\begin{eqnarray}\label{eq:S7}
\langle m\uparrow|H|n\downarrow\rangle = -\langle n\uparrow|H|m\downarrow\rangle\nonumber\\
\langle n\uparrow|H|n\downarrow\rangle = \langle n\downarrow|H|n\uparrow\rangle = 0
\end{eqnarray}
\begin{itemize}
\item{Allowed (TRS‑conserving) form.}
Only inter‑site or inter‑orbital spin‑flip terms are permitted, with the hopping amplitudes being anti-symmetric under exchange of the two states. Intra-site spin-flip elements vanish. This is the case for helical SOC in our model. Such SOC does not break TRS, and as a result, in a purely electronic model, no net spin polarization occurs\cite{sun_spontaneous_2005,guo_spin-selective_2012}.
\item{Forbidden (TRS‑breaking) form.}
Any term that allows on‑site spin flips, e.g. $t_{SO}c_{n\dagger}^{\uparrow}c_{n\downarrow}$, violates the anti-symmetry condition. Such a term effectively acts like an external Zeeman field and breaks TRS locally. It can produce spin polarization even in the absence of dissipation.
\end{itemize}

\begin{figure}[ht]
\includegraphics[width = 0.4\columnwidth]{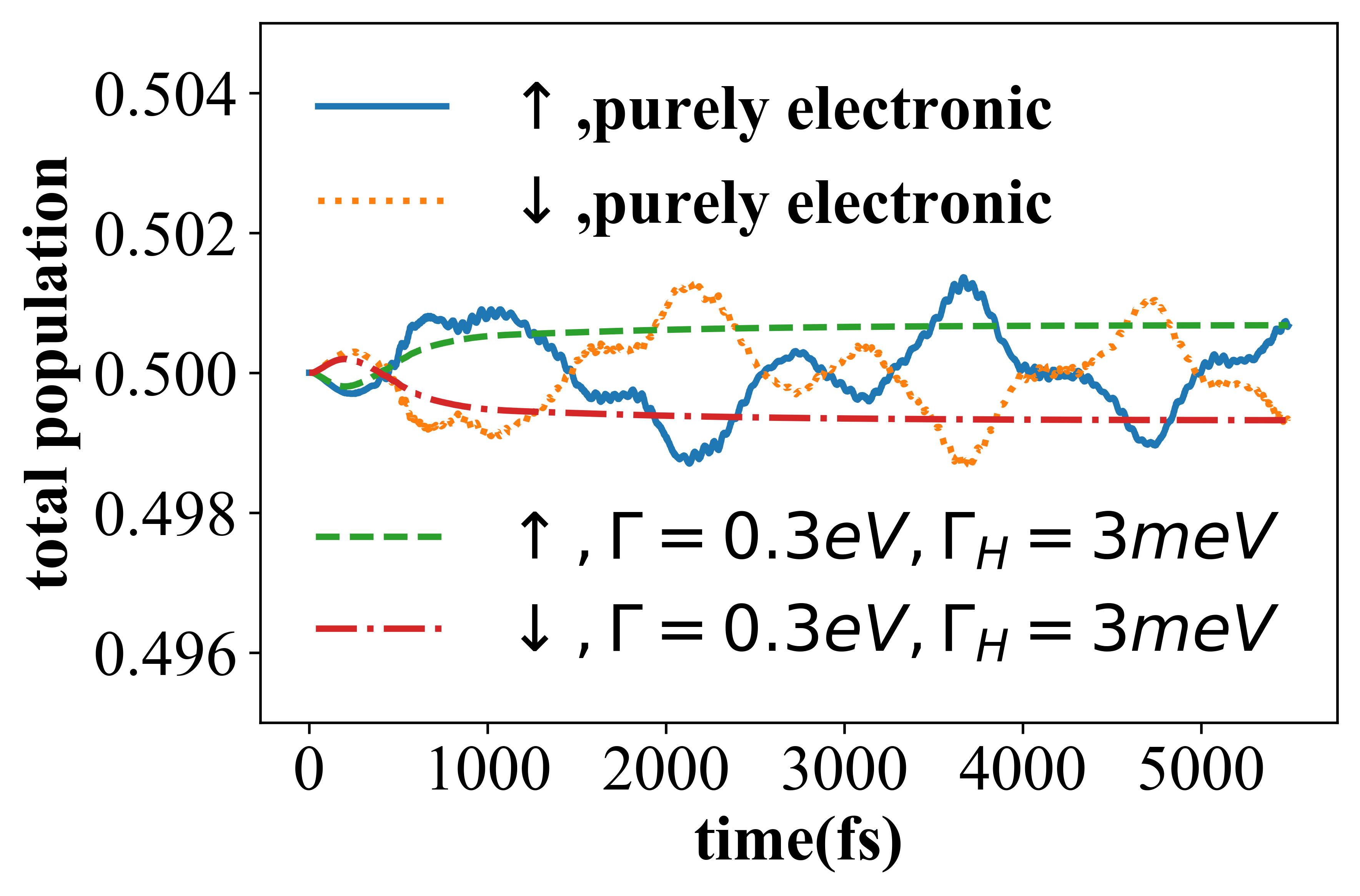}
\caption{\label{fig:e1} Total spin population as a function of time. For purely electronic model, the curves are oscillating around zero. With moderate dissipation, the population is essentially balanced and residual imbalance is less than $10^{-3}$, more than two orders of magnitude smaller than mobility polarization.}
\end{figure}

The spin‑independent Lindblad dissipation used in the main text acts equally on both spin species. Fig.~\ref{fig:e1} shows the total spin population imbalance over the entire molecule as a function of time. In the purely electronic model, the up and down populations oscillate around 50\% with almost zero net imbalance. When dissipation is included, the populations remain still almost balanced, with tiny residual imbalance being on the order of $10^{-3}$, which is more than two orders of magnitude smaller than the mobility polarization.

Consequently, the CISS effect observed in the main text is predominantly a dynamical mobility polarization, a difference in the diffusion coefficients, with negligible steady‑state spin accumulation. This is fundamentally different from scenarios where TRS-violating terms (e.g., external Zeeman field) are present. In those cases, the symmetry is broken between up and down populations, leading to a finite population polarization (i.e., a net spin density). The latter may exhibit an opposite temperature dependence (since TRS can be broken due to spin-phonon coupling)), which is beyond the scope of the present work.

\setcounter{equation}{0}
\setcounter{figure}{0}
\section{Comparison With Experimental Observations}\label{ssec:F}
In many CISS experiments, the spin selectivity is inferred from the difference in total current or conductance when the magnetization of a ferromagnetic contact is reversed\cite{naaman_chiral_2019}. Importantly, measuring the total current avoids the complications of spin-resolved measurements, which are more challenging and cannot unambiguously distinguish mobility and population contributions. Such experiments therefore probe precisely the quantity that our physical picture links to mobility polarization.

Crucially, if the observed spin selectivity were merely a population effect, i.e. spin accumulation without mobility polarization, the total current would not depend on the initial spin injection. The fact that experiments with ferromagnetic contacts have observed a dependence of the total current on the magnetization direction\cite{naaman_chiral_2019,naaman_chiral_2020} therefore indicates that the effect originates from spin-resolved transport coefficients. This is consistent with the mobility polarization predicted by our “one-enhanced, one-suppressed” bifurcation phenomenon, where the two spin channels have intrinsically different mobility.

Our results are also consistent with the observation that single-stranded DNA and achiral structures\cite{gohler_spin_2011} exhibit no CISS effect, precisely as our theoretical picture predicts.

Finally, the monotonic decrease of the mobility polarization with increasing dissipation strength $\Gamma_H$ (Fig. 4(c) of the main text) is consistent with experimental reports of decreasing CISS polarization with increasing temperature\cite{qian_chiral_2022,das_spin-induced_2023}, with predictions from spin-independent dissipation models\cite{Barroso_spin-dependent_2022} and with analysis from Ref.~\cite{liu_dynamical_2025}.

\setcounter{equation}{0}
\setcounter{figure}{0}
\section{MSD Linear‑Fit Procedure for Extracting Diffusion Coefficients}\label{ssec:G}
To extract the diffusion coefficients $D_{\uparrow}$ and $D_{\downarrow}$ from the spin‑resolved mean‑square displacement, we first identify the linear propagation regime of the wave packet. In this regime, the MSD is fitted to a linear function using least squares method. The fit is accepted only when the coefficient of determination satisfies $R^2>0.999$, ensuring excellent linearity. To minimize numerical errors, we choose the time interval as long as possible within the linear regime to get error bars as small as possible (less than $2.5*10^{-4}fs^{-1}$ for preferred spin and $2*10^{-4}fs^{-1}$ for disfavored spin), and avoiding the initial launch stage and the saturation stage.

\setcounter{equation}{0}
\setcounter{figure}{0}
\section{Time‑ and Space‑Resolved Spin Dynamics and Total Charge Transport Dynamics}\label{ssec:H}
\begin{figure}[ht]
\includegraphics[width = 0.7\columnwidth]{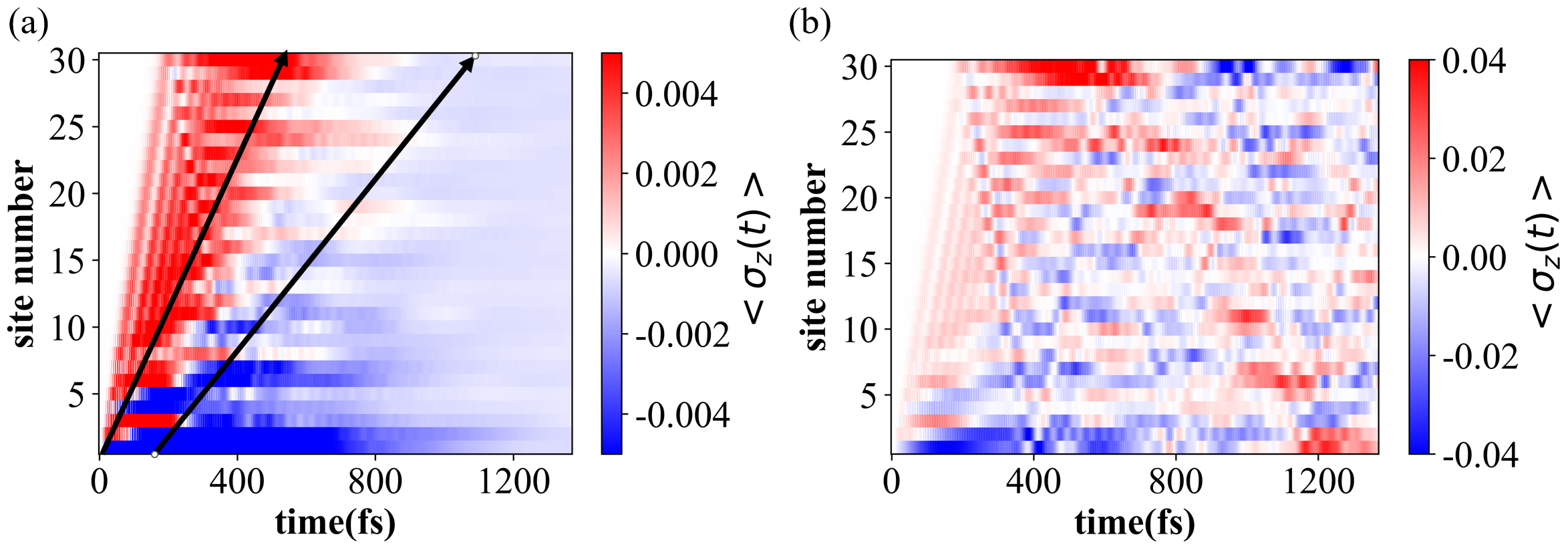}
\caption{\label{fig:h1} Two‑dimensional color maps of the spin‑dependent population on strand 2 for right-handed helix dsDNA (a) with $\Gamma=0.3eV,\Gamma_H=3meV$, (b) for purely electronic model.}
\end{figure}
Fig.~\ref{fig:h1}(a) shows two‑dimensional color maps of the spin‑dependent population on strand 2 of right-handed helix dsDNA as functions of time and site index. The donor is at site 0, the dsDNA chain spans sites 1 to N (N=30), and the acceptor is at site N+1. The maps clearly illustrate that the down‑spin wave packet lags behind the up‑spin packet throughout the propagation, reflecting the lower mobility of the down‑spin channel. In the propagation stage, the wave-packet propagation is linear with time, and the corresponding time interval (approximately 400–1000 fs, which is on the order of $N/t$ with $t=120\mathrm{meV}$ and $N=30$) matches the linear regime identified in Fig. 2(a) of the main text. In the purely electronic model (Fig.~\ref{fig:h1}(b)), the two spin populations cross periodically, resulting in zero time‑averaged polarization.
\begin{figure}[ht]
\includegraphics[width = 0.4\columnwidth]{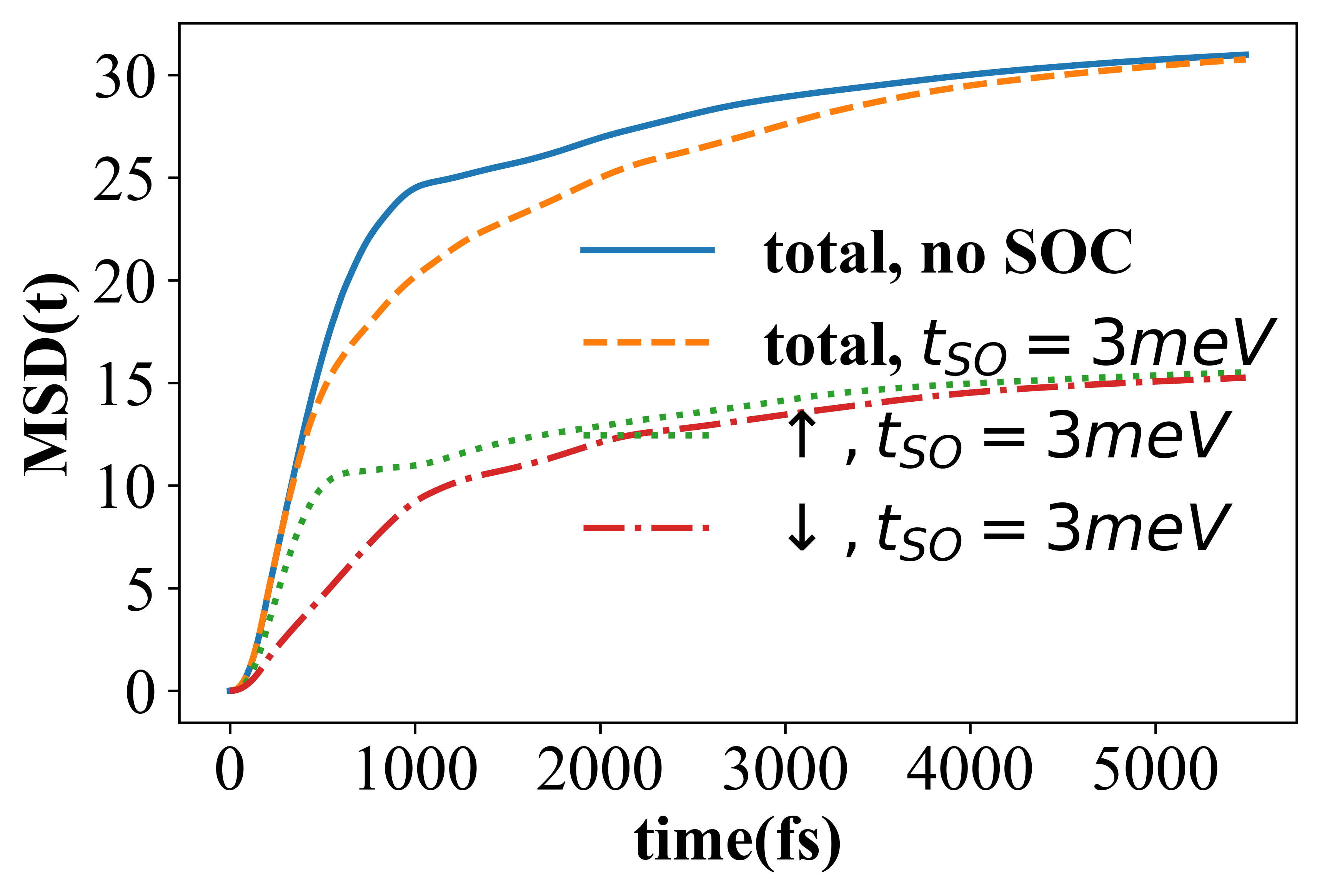}
\caption{\label{fig:h2} Total charge transport MSD (summed over up and down spins) as a function of time for the dissipative model with or without SOC. The corresponding spin-resolved MSD curves for the dissipative model with SOC are also shown.}
\end{figure}

Fig.~\ref{fig:h2} shows the total charge transport MSD (summed over both spins) for $\Gamma=0.5eV,\Gamma_H=0.6meV$. The total charge transport exhibits two distinct linear regimes. The first (faster) segment is dominated by the up‑spin propagation, which reaches the acceptor earlier. The second (slower) segment mainly reflects the down‑spin propagation, since the change of up‑spin component is gradual and its contribution is small. This observation directly corresponds to the spin‑resolved MSD shown in Fig. 2(a) of the main text, where the up‑spin and down‑spin curves are plotted separately. The figure also demonstrates the influence of SOC on the total charge transport dynamics.

\setcounter{equation}{0}
\setcounter{figure}{0}
\section{Robustness to Initial Wave‑Packet Preparation}\label{ssec:I}
\begin{figure}[ht]
\includegraphics[width = 0.4\columnwidth]{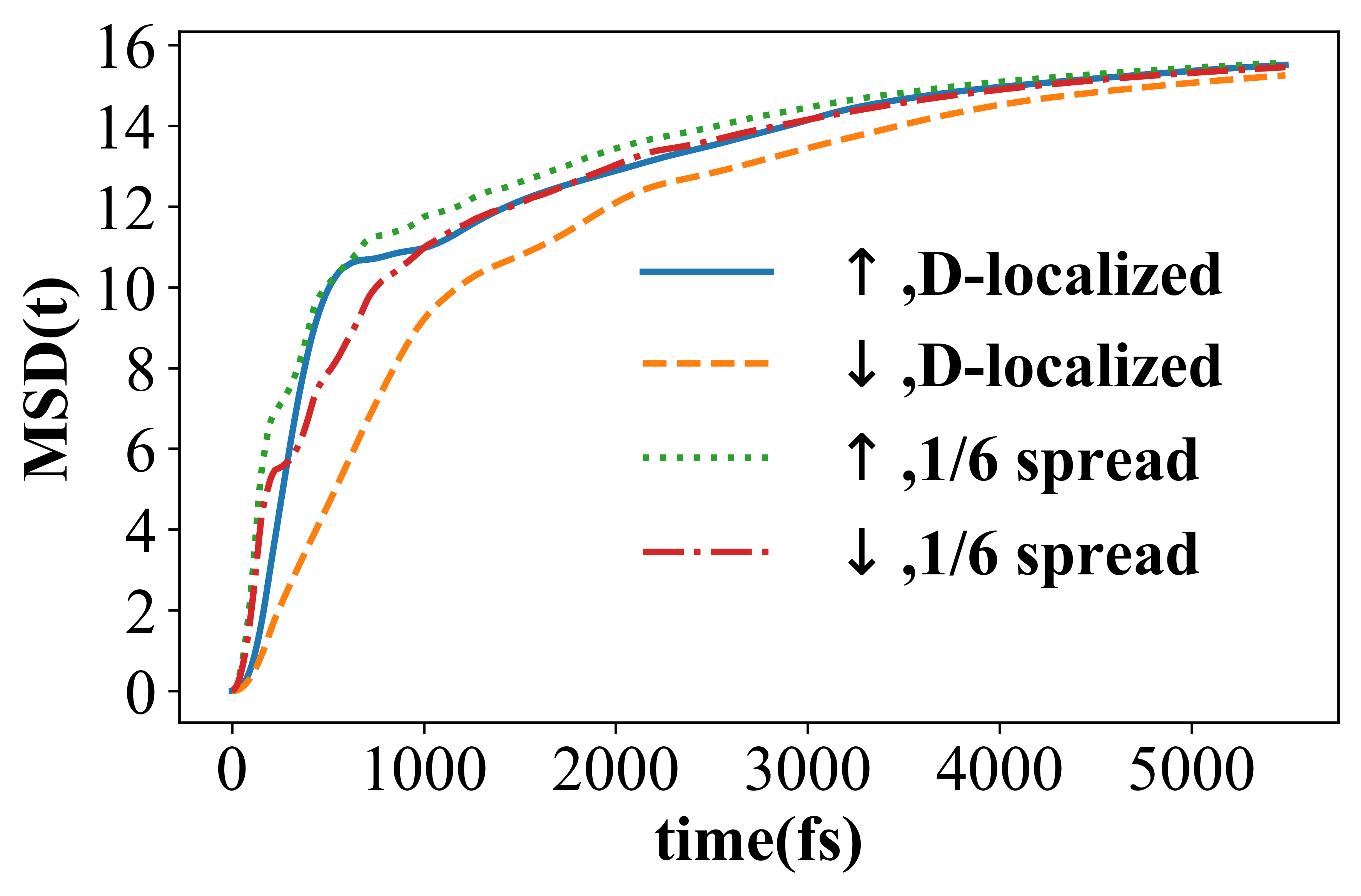}
\caption{\label{fig:i1} Spin‑resolved MSD for two initial preparations: fully localized on the donor and distributed equally among the donor and the first site of each strand (1/6 population each).}
\end{figure}
The fully localized donor initial state used in the main text is natural for the typical photo-excitation donor–acceptor experiments. This initial condition has been employed in both theoretical simulations and experimental characterizations of donor–acceptor systems\cite{balanikas_controlling_2025,athanasopoulos_ultrafast_2017}. To test the sensitivity of our results to the initial preparation, we also consider an alternative initial condition where the wave packet is initially distributed equally among the donor and the first site of each strand (1/6 population each). In this case, the spin-resolved MSD curves no longer exhibit a well-defined linear regime, which prevents a reliable extraction of the diffusion coefficients. This confirms that the fully localized donor initial state used in the main text is not only physically motivated but also essential for a well-defined mobility polarization. However, while the MSD curves may vary with the initial preparation, the qualitative spin polarization persists, supporting the robustness of the underlying theoretical picture.


In summary, the theoretical picture we have established, the “one-enhanced, one-suppressed” bifurcation phenomenon and its analytical origin, is universal and robust, remaining qualitatively unchanged.

\setcounter{equation}{0}
\setcounter{figure}{0}
\section{Generality of The ‘One-Enhanced, One-Suppressed’ Phenomenon Beyond DNA}\label{ssec:J}
To verify the generality of the “one-enhanced, one-suppressed” bifurcation phenomenon beyond the dsDNA model, we perform control calculations on a protein-like helical system. The model shares the same donor–helix–acceptor architecture but differs in two respects. It is single-stranded (no inter-strand coupling)\cite{mishra_spin-dependent_2013}, but includes long-range intra-chain interactions beyond nearest-neighbor hopping to capture the extended nature of protein electronic couplings\cite{guo_spin-dependent_2014}. The Hamiltonian follows similar forms as Eq. 1 of the main text and is adapted from Ref.~\cite{guo_spin-dependent_2014,zhang_dynamical_2025}.
\begin{figure}[ht]
\includegraphics[width = 0.4\columnwidth]{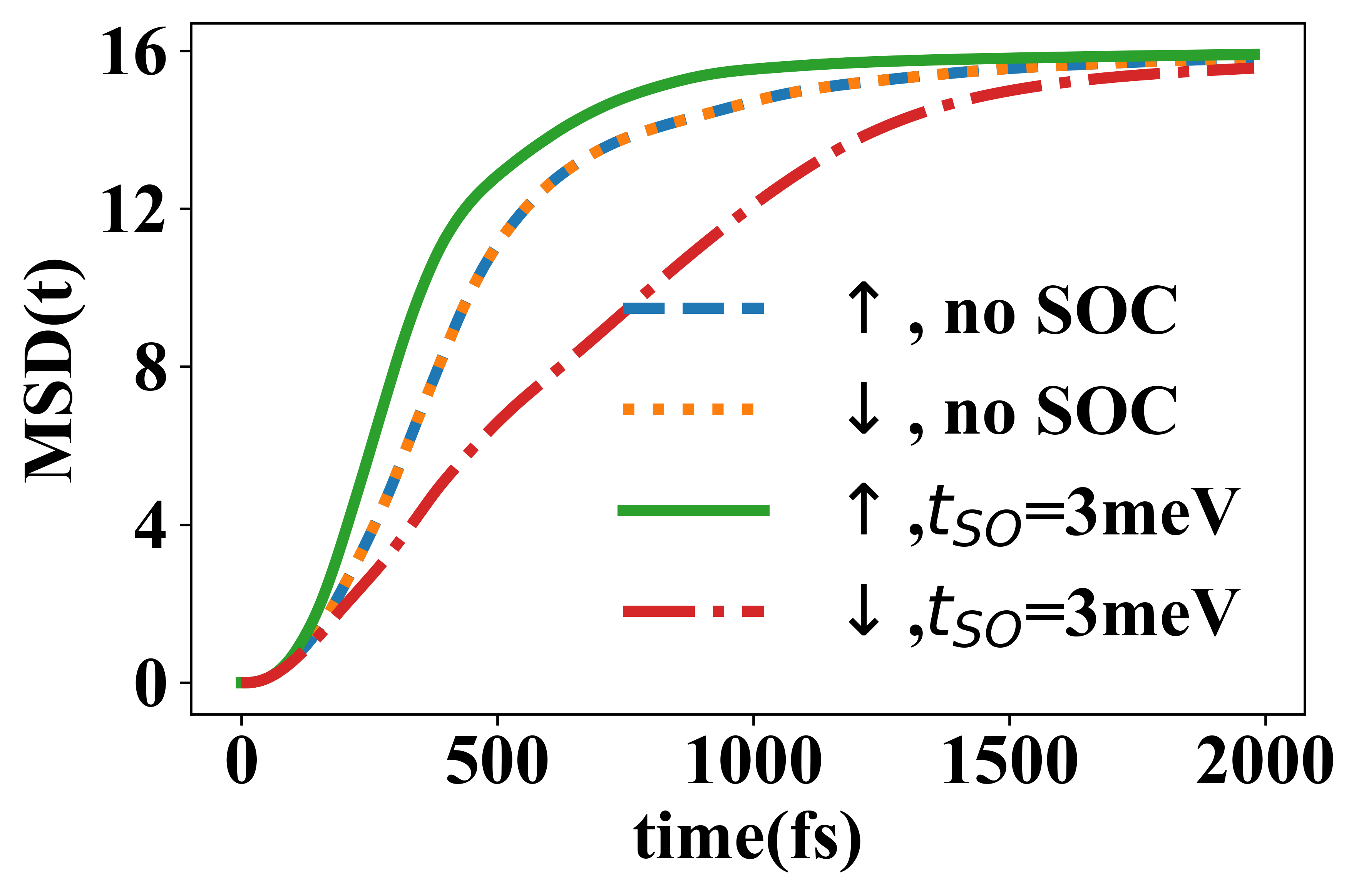}
\caption{\label{fig:j1} Verification of the “one-enhanced, one-suppressed” phenomenon in a protein-like helical model. Spin-resolved MSD dynamics for the protein-like model with and without SOC, shows the same bifurcation as in the dsDNA model.}
\end{figure}

Fig.~\ref{fig:j1} shows the spin-resolved MSD dynamics for the protein-like model under the same weak SOC strength ($t_{\mathrm{SO}}=3meV$). The two spin channels exhibit a clear bifurcation. One is enhanced and the other is suppressed relative to the SOC-free baseline, confirming the “one-enhanced, one-suppressed” pattern.

This result demonstrates that the bifurcation phenomenon is not specific to the dsDNA model but persists in qualitatively different helical systems. This confirms that the bifurcation phenomenon and the CISS effect arise as general consequences of helicity, multi-path interference, weak SOC, and dissipation, rather than a DNA-specific effect.

\end{document}